\documentclass[
]{ceurart}

\usepackage{listings}
\usepackage{booktabs}
\usepackage{graphicx}
\usepackage{booktabs}
\usepackage{adjustbox}
\usepackage{graphicx}
\usepackage{subcaption} 
\usepackage{multirow}
\usepackage{placeins}
\usepackage{tabularx}
\begin{document}

\copyrightyear{2025}
\copyrightclause{Copyright for this paper by its authors.
  Use permitted under Creative Commons License Attribution 4.0
  International (CC BY 4.0).}

\conference{13th International Workshop on News Recommendation and Analytics (INRA), September 22--26, 2025, Prague, Czech Republic}

\title{Leveraging Media Frames to Improve Normative Diversity in News Recommendations}

% \tnotetext[1]{You can use this document as the template for preparing your
  % publication. We recommend using the latest version of the ceurart style.}

\author[1]{Sourabh Dattawad}[
    orcid=0000-0002-4642-0743,
    email=dattawsh@ims.uni-stuttgart.de,
    url=https://sourabhdattawad.github.io/sourabhdattawad/,
]
\cormark[1]
\address[1]{Institute for Natural Language Processing, University of Stuttgart, Germany}

\author[1]{Agnese Daffara}[
    orcid=0009-0009-6706-0095,
    email=agnese.daffara@ims.uni-stuttgart.de, 
    url={https://scholar.google.com/citations?user=JdOLC-IAAAAJ},                
]

\author[2]{Tanise Ceron}[
    orcid=0009-0002-4845-2789,
    email=tanise.ceron@unibocconi.it, 
    url=https://tceron.github.io,        
]
\address[2]{Department of Computing Sciences, Bocconi University, Italy}

\cortext[1]{Corresponding author.}
% \fntext[1]{These authors contributed equally.}

\begin{abstract}
Click-based news recommender systems suggest users content that aligns with their existing history, limiting the diversity of articles they encounter. Recent advances in aspect-based diversification -- adding features such as sentiments or news categories (e.g. world, politics) -- have made progress toward diversifying recommendations in terms of perspectives. However, these approaches often overlook the role of news framing, which shapes how stories are told by emphasizing specific angles or interpretations. In this paper, we treat media frames as a controllable aspect within the recommendation pipeline. By selecting articles based on a diversity of frames, our approach emphasizes varied narrative angles and broadens the interpretive space recommended to users. In addition to introducing frame-based diversification method, our work is the first to assess the impact of a news recommender system that integrates frame diversity using normative diversity metrics: representation, calibration, and activation. Our experiments based on media frame diversification show an improvement in exposure to previously unclicked frames up to 50\%. This is important because repeated exposure to the same frames can reinforce existing biases or narrow interpretations, whereas introducing novel frames broadens users’ understanding of issues and perspectives.
The method also enhances diversification across categorical and sentiment levels, thereby demonstrating that framing acts as a strong control lever for enhancing normative diversity.
\end{abstract}


\maketitle

\section{Introduction}

Recent studies have highlighted the importance of diversifying recommendations in news recommendation systems to expose users to a broader range of perspectives \cite{reuver-etal-2021-nlp}. For example, news recommenders with well-designed normative goals ensure exposure to multiple viewpoints, supporting deliberation, and amplifying underrepresented voices \citep{helberger2021democratic}.
%Such diversification can foster democratic participation in public discourse and help mitigate the creation of closed informational spaces such as filter bubbles \citep{filterbubble} and echo chambers \citep{Sunstein2007}, where ideas fail to circulate freely.%
Such diversification is expected to foster democratic participation in public discourse, although empirical evidence remains limited. Recent work highlights that normatively motivated diversification can, under certain conditions, support tolerance and participation, even though its effects are not yet fully established due to the difficulty in evaluation \citep{Mattis12032025}. 
%Similarly, while Pariser’s early work on filter bubbles \citep{filterbubble} introduced the concept anecdotally, subsequent studies provide more nuanced perspectives on the extent and mechanisms of personalization effects \citep{hartmann2025systematicreviewechochamber, doi:10.1073/pnas.2318127122}. Importantly, countering filter bubbles is only one motivation for diversification;  
Existing approaches have made progress in incorporating different perspectives in recommendation models, operationalizing them as "aspects" such as news categories, emotions, and sentiments~\cite{wu-etal-2020-sentirec, Babanejad2019LeveragingEF, qi2022profairrecproviderfairnessawarenews}. However, they have not considered how news stories are narrated or framed. To address this challenge, we propose diversifying perspectives in news recommendations through framing in news coverage. While a previous study by ~\citet{mulder2021operationalizing} already explored frame-based news re-ranking, our work is, to the best of our knowledge, the first to apply the full set of media frames from the Media Frames Corpus (MFC) ~\citep{Card2015TheMF} to the core challenge of personalized news recommendation. Besides that, previous works often do not address how encoding different aspects enhances
normative diversity, a social-scientific concept reflecting the norms and values of news organizations—such as those emphasized in different models of democracy: personal development and autonomy (liberal model), active citizenship (participatory model), access to different viewpoints (deliberative model), access to underrepresented voices (critical model) \citep{helberger2021democratic}. 

In the widespread definition by \citet{entman1993}, framing means selecting some parts of reality by making them more salient, with four functions: (i) defining problems, (ii) diagnosing causes, (iii) making judgments, and (iv) suggesting solutions. Framing has received much attention in Media Studies and NLP, as it can shape the dissemination of information and ultimately influence people’s beliefs \citep{lecheler2015effects}. For example, it can serve as a powerful tool for agenda-setting and manipulation strategies, because authors can use it to intentionally present some aspects while omitting others \citep{tsur-etal-2015-frame, field-etal-2018-framing}. Given its connection to political ideologies \citep{alashri2015animates} and its multifaceted nature, which encompasses different semantic and pragmatic levels \citep{otmakhova2024media}, framing may prove particularly well-suited for analyzing perspectives in news articles and can effectively serve the purpose of diversifying news recommendations. 

We identify two open research questions: 

\begin{enumerate}
    \item Do media frames enhance normative diversity in news recommendations?
    \item To what extent can we tune recommender systems with media frames to improve the normative quality of news recommendations? 
\end{enumerate}

We implement a frames detection model trained on the MFC \citep{Card2015TheMF}.
We then integrate these frames as an aspect into the existing MANNeR framework, a news recommendation algorithm that supports aspect-based diversification \citep{iana2024trainonceuseflexibly}. The MANNeR framework enables flexible, aspect-aware news recommendation by modularly combining multiple specialized encoders. This design supports both personalization and diversification by allowing targeted control over content dimensions like frames, categories, or sentiment. It is particularly well-suited for our setting due to its efficiency in training and the ease with which it integrates new aspects, eliminating the need for full retraining.
To evaluate the system's ability to reflect diverse frames, we adapt the RADio* metrics, which capture various dimensions of normative diversity \citep{Vrijenhoek}. Additionally, we leverage MANNeR’s diversification strategy based on linear score aggregation to fine-tune the amount of frame diversity according to recommendation quality or user needs.

We train and evaluate our models in a multilingual setting using two datasets: the News Portal Recommendations (NPR) dataset in Portuguese \citep{Lucas2023NPRAN}, and the Ekstra Bladet News Recommendation Dataset (EB-NeRD) in Danish \citep{Kruse_2024}. Our experiments demonstrate that this approach significantly increases exposure to diverse and novel perspectives that users had not previously encountered. Interestingly, we also observe that framing inherently captures both the sentiment and the category of news articles. This enables us to achieve sentiment- and category-based diversification implicitly to an extent, without needing to explicitly encode these features during training. We demonstrate that framing can serve as a powerful lever for steering normative diversity in news recommendation systems, enabling the tuning of aspects such as news category, sentiment, and frames to achieve the desired outcome.

\section{Related Work}

\subsection{Diverse-Aware News Recommenders}

The workflow of a news recommender system involves recalling a subset of articles from a large corpus and then ranking them based on user interests. Central to this process is the creation of a user embedding that accurately captures their interests, as well as a news embedding that captures the article's core semantic meaning. For instance, models like NAML \cite{wu2019neuralnewsrecommendationattentive} approaches this by using CNNs to encode the news content a user reads and GRUs to learn a sequential representation of their interests. Subsequent research has built upon this, with models like NRMS \cite{wu-etal-2019-neural-news} and HieRec \cite{qi2021hierechierarchicalusermodeling} introducing attention mechanisms and hierarchical structures to create even richer user embeddings for the ranking task. 

To tackle the impact excessive personalization and of filter bubbles -- where users are repeatedly exposed to similar content -- diversity has become a central focus in recommender systems. Beyond algorithmic concerns, user studies have shown that a lack of diversity can lead to frustration \citep{10.1145/1454008.1454030, 10.1145/1835449.1835486}, and that diversity plays a key role in user acceptance, especially when different models perform similarly in terms of accuracy \citep{10.1145/2645710.2645737, item_af1f365bbc7d4ee682c3bc20d2025fb1}. \citet{10.1145/1864708.1864717} highlight that diversity increases user confidence in decision-making, while \citet{10.1145/3437963.3441786} show it helps elicit user preferences and reduces popularity bias.
With respect to diverse news recommendation—where the goal is to recommend content dissimilar to a user’s history in terms of sentiment, emotion, categories, or sources—several approaches have emerged. SentiRec enhances sentiment diversity by explicitly optimizing for a range of sentiment polarities, thus preventing overexposure to emotionally homogeneous content \citep{wu-etal-2020-sentirec}. Emotion-aware models leverage emotional representations of users and articles to recommend affectively novel content, fostering emotional diversity in consumed news \citep{Babanejad2019LeveragingEF}. EmoRec incorporates multi-level emotional signals, capturing both coarse and fine-grained affect to provide recommendations that contrast with prior emotional experiences \citep{10.1145/3511047.3536414}. ProFairRec addresses source-side diversity by promoting provider fairness through debiased embeddings and fairness-aware learning objectives, ensuring exposure to underrepresented news outlets \citep{qi2022profairrecproviderfairnessawarenews}. Graph-based topic nudging exposes users to novel topical areas by expanding and nudging topic representations via knowledge graphs, increasing topic-level dissimilarity from prior consumption \citep{vercoutere-2025-graph}. {D\textsuperscript{2}RL} combines Determinantal Point Processes with actor-critic reinforcement learning to reward both diversity and relevance, promoting dissimilarity in content ranking over time \citep{liu2019diversitypromotingdeepreinforcementlearning}. Finally, MANNeR introduces multi-aspect diversification by simultaneously optimizing for topical variety and sentiment diversity \citep{iana2024trainonceuseflexibly}. 

\subsection{Diversity Metrics for Recommender Systems}
Diversity in recommender systems can be broadly categorized into two types: descriptive (general-purpose) and normative. Overall, descriptive diversity provides computational tools to increase variety, while normative diversity aligns recommender systems with ethical and democratic responsibilities \citep{Vrijenhoek2021}. 
The former aims to reduce redundancy in recommendation lists using computational measures of dissimilarity. \citet{10.1145/1060745.1060754} introduce Intra-List Similarity (ILS) and Intra-List Distance as foundational metrics. Other measures include relative diversity, expected intra-list diversity (which incorporates ranking sensitivity), aggregate diversity (across user lists), and the inverse Gini index to capture distributional inequality \citep{bradley2001, 10.1145/2043932.2043955, RePEc:net:wpaper:0710}. However, as \citet{10.1145/3511047.3538030} note, these are often single-number metrics that only partially address broader news recommendation challenges.

Normative diversity extends beyond computational dissimilarity to promote societal goals such as inclusion of voices and opinions, having the objective to strengthen democratic values. \citet{Helberger01012018} propose a cooperative responsibility framework, emphasizing the shared role of platforms, users, and institutions in realizing public values. \citet{Vrijenhoek2021} introduce the DART metrics -- calibration, fragmentation, representation, activation, and alternative voices -- to capture normative diversity dimensions. While insightful, these lack rank-awareness and uniformity, issues addressed in the RADIo* framework \citep{Vrijenhoek}, which offers rank-sensitive and consistent metrics for evaluating exposure diversity in recommendation lists. See Section \ref{eval} for the precise definitions and computation details of these normative metrics.

\subsection{Frames Detection and Application in News Recommendation}
Framing categories can be recognized through specific textual and visual devices (e.g. figures, videos)
%\seb{what are 'visual devices'?}
and can be either inducted from the text or established \textit{a priori} \cite{de2005news}. The task of frame detection was first explored in unsupervised settings by treating frames as special types of subtopics \citep{tsur-etal-2015-frame, nguyen-etal-2015-tea, Boydstun2013IdentifyingMF, roberts2014structural, ajjour2019modeling}. It has since been addressed using supervised or semi-supervised classification techniques, including neural networks \citep{naderi-hirst-2017-classifying}, fine-tuning of pre-trained models \citep{kwak2020systematicmediaframeanalysis}, and zero-shot prompting \cite{gilardi2023chatgpt,daffara2025generalizability}. 

A widely used inventory of generic frames is the one introduced by \citet{Boydstun2013IdentifyingMF} and adopted by \citet{Card2015TheMF} in the Media Frame Corpus (MFC), which comprises 15 categories, such as \textit{Politic}, \textit{Morality}, and \textit{Fairness and Equality} (see Table \ref{tab:frames_examples} for a complete list with examples).
This annotation framework has been applied extensively in subsequent studies \cite{Khanehzar2019ModelingPF, khanehzar2021framing, mendelsohn-etal-2021-modeling, mulder2021operationalizing, gilardi2023chatgpt, piskorski2023semeval}. SemEval-2023 Task 3 adopted it for a sub-task on framing detection in online news within a multilingual context--comprising 9 different languages--where all participating teams employed transformer-based models \citep{piskorski2023semeval}. Although the label set was originally developed from 3 policy issues in the U.S. context, and relying on these coarse-grained categories may lead to some loss of specificity across datasets, the MFC frames have been demonstrated to be generalizable to new contexts and policy issues \cite{daffara2025generalizability}. We therefore adopt this set in our study.

\citet{mulder2021operationalizing} 
are the first to apply framing to diverse recommendation. They propose a re-ranking method for news recommendation lists based on the four functions identified by \citet{entman1993}: (i) Problem Definition, (ii) Causal Attribution, (iii) Moral Evaluation, and (iv) Treatment Recommendation.
Specifically, they implement different extraction methods tailored to each constituent: for Problem Definition, they use LDA topic modeling to infer the central issue presented in each article; for Causal Attribution and Moral Evaluation, they apply the IBM Watson Natural Language Processing API to classify news content into topic-related taxonomies and extract sentiment; for Treatment Recommendation, they extract candidate sentences and classify them using the same taxonomy-based classifier. These identified frame functions are then used to measure diversity across articles using different distance metrics, including Kullback-Leibler divergence for LDA-derived topics and a weighted Jaccard index for the taxonomy-based categories. This allows them to re-rank news articles to maximize framing diversity in the recommendation list. While their work focuses on extracting and balancing these four framing functions, our approach shifts attention to the 15 MFC categories, applying them in a topic-generic setting. Rather than focusing on framing constituents, we aim to capture the type of frame conveyed by each article across multiple news categories while leveraging normative diversity. 
\begin{table}[!h]
\centering
\scriptsize
\begin{tabular}{@{}p{0.27\linewidth} p{0.70\linewidth}@{}}
\toprule
\textbf{Frame} & \textbf{Example} \\ \midrule
Capacity and Resources & \textit{Immigration debate: Illegals take jobs from Americans} \\
Crime and Punishment & \textit{Two charged in deaths of illegal immigrants in truck} \\
Cultural Identity & \textit{Ethnic shift: Immigration—an Irish enclave learns a new language; Mexican immigrants boost a growing Latino population} \\
Economic & \textit{Society makes no-interest loans to New York's immigrants} \\
Fairness and Equality & \textit{Strict immigration law unfairly targets Hispanics} \\
External Regulation and Reputation & \textit{‘International village’ gets hostile reception} \\
Health and Safety & \textit{Colombian drug violence leads to exodus} \\
Legality, Constitutionality and Jurisprudence & \textit{House approves bill to abolish INS; The Senate will begin work next week on its own measure dealing with the immigration agency} \\
Morality & \textit{County's undocumented workers say they aren't here to ‘steal’} \\
Other & \textit{U.S. under pressure to carry bigger load} \\
Policy Prescription and Evaluation & \textit{President Donald Trump stalls on promise to eliminate J-1 visa program} \\
Political & \textit{Following Trump voter fraud allegations, claim that 5.7 million non-citizens voted is wrong} \\
Public Opinion & \textit{Immigration: Political evangelicals feel push to take sides} \\
Quality of Life & \textit{Big money, cheap labor} \\
Security and Defense & \textit{Decision on refugees overdue; U.S. officials must loosen immigration restrictions} \\
\bottomrule
\end{tabular}
\caption{The 15 frames and their examples extracted from the MFC headlines on migration.}
\label{tab:frames_examples}
\end{table}

\section{Methodology}

The process of incorporating media frame awareness into a news recommendation system involves two main stages. In the first stage, we employ a validated classification model to automatically assign a primary media frame to each article. Following \citet{Card2015TheMF}, we define the primary frame as the single, most prominent frame in the text. While articles can contain multiple underlying frames in their title and body, we only focus on the primary one as it has been shown to be the most influential in shaping a reader's interpretation and serves as a powerful feature for frame-aware recommendation.

Second, we extend the modular MANNeR recommendation framework by creating a dedicated module that represents articles based on their frame similarity. This allows us to combine content-based personalization with a new, frame-based dimension, enabling explicit control over the diversity of recommended news. Lastly, we evaluate the frame-aware news recommenders with normative metrics adapted from RADIo*.

\subsection{Media Frames Detection}

In our study, we adopt the methodology established by prior work~\cite{daffara2025generalizability} to identify the dominant media frame for each article in our corpus. Specifically, we use a pre-trained \texttt{XLM-RoBERTa-base} model that has been fine-tuned on the MFC dataset~\cite{Card2015TheMF}. The model assigns a probability distribution over all the 15 frame labels for each article, and we select the one with the highest probability as the primary frame. Table \ref{tab:frames_examples} illustrates examples of the frames used in the training.

Regarding the evaluation of this classifier, on the MFC test set, the fine-tuned model achieved an accuracy of 67\% and an F1 score of 68\%. Frame-level results indicate high predictive performance on common categories, such as Political (F1: 79.85\%) and Capacity and Resources (F1: 70.33\%). To assess the model’s robustness in out-of-distribution settings, it is evaluated on \textit{FrameNews-PT}, a dataset of 300 manually annotated Portuguese news articles using the same frame schema \citep{daffara2025generalizability}. On this dataset, the model achieved a reduced accuracy of 53\% and an F1 score of 0.48, though it still outperformed the baseline (accuracy: 23\%). Notably, performance was particularly weak on abstract or infrequent frames such as \textit{Morality}, \textit{Fairness and Equality}, and \textit{Quality of Life}, mirroring categories with low inter-annotator agreement (Krippendorff’s $\alpha = 0.78$ overall). These results highlight known challenges in generalizing U.S.-centric frame classifiers across languages and cultural contexts—challenges we account for in our own analysis.

\subsection{Integration of Media Frames into Recommendation System}

The MANNeR framework introduces a modular approach to news recommendation, enabling the flexible integration of various aspects beyond pure content similarity. It achieves this by training specialized News Encoders (NEs) for different aspects (which the authors call A-module), which are then linearly combined at inference time. Building on this, we extend the MANNeR framework to incorporate media frames as an aspect for recommendation, while disabling other aspects originally supported by the framework, such as topics and sentiments. 

The content-based personalization module (CR-Module) remains unchanged. Both the candidate and the clicked news articles are encoded using a dedicated News Encoder (NE). The candidate embedding $n_c$ is compared to each clicked news embedding $n_{u_i}$ using a dot product, and the resulting similarity scores are mean-pooled to compute the overall content relevance score:
\begin{equation}
s_{\text{CR}}(n_c, u) = \frac{1}{N} \sum_{i=1}^{N} n_c \cdot n_{u_i}
\end{equation}

The CR-Module is trained by fine-tuning the underlying Pretrained Language Model (PLM) using Supervised Contrastive Learning (SCL). For each user, clicked news items are treated as positive samples, while non-clicked items—selected via negative sampling—serve as negative examples. The training objective encourages the model to draw the candidate embedding closer to clicked news and push it away from non-clicked ones, thereby structuring the embedding space according to user relevance.

We introduce a dedicated A-Module for media frames while the original MANNeR framework has one A-module for sentiment and one for topics.
Following the MANNeR methodology, the Frame-based A-Module, which we call Frame-Module from now on, is trained to create a specialized news representation space. This is achieved by fine-tuning a separate copy of the initial PLM using a SCL. The objective is to group news articles with the same frame label closer together in the embedding space while simultaneously pushing those with different frames further apart. As with other A-Modules, this module encodes news similarity specifically with respect to frames, independent of user preferences.

At inference time, we leverage the specialized NEs from both the CR-Module and the Frame-Module. For a given candidate news article ($n_c$) and the user's click history ($H = \{n_{u_1}, n_{u_2}, ..., n_{u_N}\}$), we compute two separate similarity scores. The content similarity ($s_{\text{CR}}$) is calculated as the mean-pooled dot product of the candidate's and the clicked news's embeddings from the CR-Module. Similarly, the frame similarity ($s_{\text{frame}}$) is computed using the embeddings from the Frame-Module.

The final ranking score ($s_{\text{final}}$) for a candidate news item ($n_c$) for a user ($u$) is a linear aggregation of z-score normalized scores:
\begin{equation}
\label{eq:ranking_score}
s_{\text{final}}(n_c, u) = s_{\text{CR}}(n_c, H) + \lambda_{\text{frame}} s_{\text{frame}}(n_c, H)
\end{equation}

In this equation, the hyperparameter $\lambda_{\text{frame}}$ controls the influence of the media frame aspect on the final recommendation. Setting $\lambda_{\text{frame}} > 0$ promotes personalization towards news with similar frames to what the user has previously engaged with, while $\lambda_{\text{frame}} < 0$ will encourage diversification by recommending news with different frames.

\section{Evaluation}

Our baseline is the CR-module alone as it does not diversify content based on aspects. Moreover, we evaluate how the frame-aware news recommenders perform at different values of $\lambda$, ranging from -1 to 1. Since the underlying scores are z-score normalized prior to the linear combination, their scales are made comparable, which justifies the use of this symmetric range for $\lambda$. We evaluate all systems in the following metrics: 

\paragraph{Descriptive evaluation} We evaluate our news recommendation systems using standard ranking metrics: AUC, MRR, and nDCG@k (with \(k \in \{5, 10\}\)), which respectively measure the ability to distinguish relevant from non-relevant items, how early relevant items appear in the ranking, and the overall quality of the ranking with greater emphasis on higher-ranked relevant items. 

\paragraph{Normative evaluation}
To assess normative diversity, we adopt the metrics defined in RADio*~\citep{Vrijenhoek}, modifying them to capture the framing dimension. The value for these metrics ranges from 0 to 1. Formally, we adapt the method defined in RADio* and extend it to capture frame diversity. Specifically, we define the diversity of a recommendation list $P$ with respect to a context $Q$ in terms of the distribution of media frames as:

\[
D_f^*(P, Q) = \sum_{x \in \text{Frames}} Q^*(x) \cdot f\left(\frac{P^*(x)}{Q^*(x)}\right)
\]

Here, $x \in \text{Frames}$ refers to each possible media frame (e.g., economic, morality, legality). $Q$ denotes the list of recommended articles, and $P$ denotes the context list which provides the ground truth, such as a user's reading history or a reference news corpus. $Q^*(x)$ is the proportion of frame $x$ in the recommendation list, and $P^*(x)$ is the corresponding proportion in the context list. Both the recommendation list $Q$ and the context $P$ can be made rank-aware. The function $f$ denotes a divergence metric, viz.\ Jensen-Shannon divergence (JSD). Table \ref{tab:metrics_summary_adjusted} shows the summary features used for the computation of these metrics. The following measures are used:
\begin{itemize}
    \item \textbf{Calibration} – It measures how well recommendations match a user’s historical preferences. In our setting, it compares the distribution of both news categories and frames in the recommended items with those in the user’s history.  
    \textit{Low scores} indicate strong personalization, while \textit{high scores} suggest more novel or diversified recommendations in terms of news categories and frames.

    \item \textbf{Representation} – It evaluates how closely the distribution of frames in the recommended content mirrors the overall distribution in the dataset.  
    \textit{Low scores} mean recommendations align with the dataset’s frame distribution, while \textit{high scores} indicate a deviation that may highlight over- or under-represented frames. 

    \item \textbf{Activation} – This metric measures the emotional intensity of recommended content based on the absolute sentiment score of articles. \textit{Low activation} indicates alignment with the typical tone of the platform, while \textit{ high activation} reflects the divergence of recommended content from the typical tone of the platform. This metric helps assess how framing influences emotional engagement. For computing sentiment scores of the articles, we use \texttt{XLM-T}, a multilingual sentiment analysis model \citep{barbieri-etal-2022-xlm}.

\end{itemize}
Table \ref{tab:metrics_summary_adjusted} shows the summary features used for the computation of these metrics. For representation, we adopt the dataset frame distribution as the contextual baseline, as it reflects the editorial and topical balance of the source corpus. This choice is not without limitations. A uniform distribution could alternatively be used if the objective were to maximize exposure to all frames equally, but such an approach would impose a normative stance that may not be realistic or desirable across all contexts. For example, in some domains, aiming for proportionality with the dataset better captures what users would typically encounter in the news ecosystem, whereas in others, equal representation of frames may be preferable to broaden perspectives. Similarly, while our analysis interprets higher divergence as an indicator of improved alignment or representativeness, we note that this is not a universal truth. The desirability of higher or lower divergence depends on the specific normative goals of the recommendation system (e.g., mirroring existing distributions vs. actively counterbalancing them). We therefore present our results as evidence of controllable diversity effects rather than prescribing a single ‘correct’ target distribution.

\label{eval}

\begin{table}[!h]
\centering
\begin{adjustbox}{width=1\textwidth, center}
\begin{tabular}{lccccl}
\toprule
\textbf{Metric} & \textbf{Feature} & \textbf{Context} & \textbf{Type} & \textbf{Rank-aware} & \textbf{Desired Value} \\
\midrule
Calibration (Category \& Frame) & Category, Frame & User history & Categorical & Yes & Low: reflects cats, High: diverse cats \\
Representation (Frame) & Frame class & All articles & Categorical  & No & Low: reflect frames, High: diverse frames \\
Activation (Frame) & Sentiment score & All articles & Continuous (binned) & No & Low: typical tone, High: untypical tone \\
\bottomrule
\end{tabular}
\end{adjustbox}
\caption{Summary of metrics with features, context, and interpretation. Cats stands for news categories.}
\label{tab:metrics_summary_adjusted}
\end{table}

\subsection{Datasets}

We choose to work with the datasets NPR \citep{Lucas2023NPRAN} and EB-NeRD \cite{Kruse_2024} because, among the click-history news recommendation datasets, they include the largest amount of hard news
(e.g., politics, economics, finance). Datasets with a higher proportion of hard news are more suitable for studying normative diversity, which emphasizes the influence of news on public discourse and civic engagement \citep{Lucas2023NPRAN,
vrijenhoek2023mindreflectionsminddataset}. 
\paragraph{\textbf{NPR}} It was developed by Globo media, an organization based in Brazil
\cite{Lucas2023NPRAN}. This dataset includes 1,162,802 randomly sampled users, 148,099 Portuguese news articles, and 1,402,576 impression logs. NPR was specifically designed to support research on normative diversity, making it well-suited for studying the societal impact of hard news content.
\paragraph{\textbf{EB-NeRD}} It was compiled from user behavior logs of the Danish newspaper Ekstra Bladet \cite{Kruse_2024}, EB-NeRD features over 1 million users, 37 million impression logs, 251 million interactions, and 125,000 Danish news articles. It emphasizes text-based recommendations for low-resource languages and includes a high proportion of hard news categories such as crime, politics, and economics.

\begin{table}[!h]
\centering
\scriptsize % smaller than \small
\begin{adjustbox}{width=0.6\columnwidth, center} % reduce width a bit more
\begin{tabular}{lr|lr}
\toprule
\multicolumn{2}{c|}{\textbf{NPR}} & \multicolumn{2}{c}{\textbf{EB-NeRD}} \\
\cmidrule(r){1-2} \cmidrule(l){3-4}
\textbf{Category} & \textbf{Prop. (\%)} & \textbf{Category} & \textbf{Prop. (\%)} \\
\midrule
sp        & 14.95 & nyheder       & 20.95 \\
mg        & 7.37  & sport         & 16.87 \\
rj        & 6.82  & underholdning & 16.79 \\
mundo     & 4.63  & krimi         & 14.42 \\
bahia     & 4.22  & side9         & 7.76  \\
política  & 4.12  & auto          & 7.07  \\
go        & 3.83  & forbrug       & 4.58  \\
são paulo & 3.58  & sex\_og\_samliv & 3.70  \\
economia  & 3.55  & nationen      & 2.38  \\
pe        & 2.90  & musik         & 1.87  \\
pr        & 2.89  & ferie         & 0.76  \\
pa        & 2.48  & play          & 0.62  \\
ce        & 2.25  & biler         & 0.60  \\
df        & 2.07  & penge         & 0.58  \\
to        & 1.91  & opinionen     & 0.32  \\
rs        & 1.85  & haandvaerkeren& 0.19  \\
Other     & 23.67 & Other         & 4.83  \\
\bottomrule
\end{tabular}
\end{adjustbox}
\caption{Proportion of the top categories in the NPR and EB-NeRD datasets. Note that in the NPR dataset, the two-character categories represent the acronyms of Brazilian states.}
\label{tab:npr_ebnerd_proportions_only}
\end{table}

\begin{table}[!h]
\centering
\begin{adjustbox}{width=0.75\columnwidth, center}
\begin{tabular}{lrr}
\toprule
\textbf{Frame} & \textbf{NPR-small (\%)} & \textbf{EB-NeRD-small (\%)} \\
\midrule
Economic                                      & 8.20  & 11.65 \\
Capacity and resources                        & 20.08 & 0.60  \\
Morality                                      & 9.70  & 33.48 \\
Fairness and equality                         & 2.89  & 17.28 \\
Legality, constitutionality and jurisprudence & 14.23 & 11.23 \\
Policy prescription and evaluation            & 7.35  & 5.73  \\
Crime and punishment                          & 1.66  & 8.27  \\
Security and defense                          & 0.73  & 2.74  \\
Health and safety                             & 1.39  & 4.31  \\
Quality of life                               & 27.53 & 2.58  \\
Cultural identity                             & 4.21  & 1.19  \\
Public opinion                                & 0.94  & 0.53  \\
Political                                     & 0.95  & 0.27  \\
External regulation and reputation            & 0.15  & 0.14  \\
Other                                         & 0.00  & 0.00  \\
\bottomrule
\end{tabular}
\end{adjustbox}
\caption{Proportion of each frame in the datasets.}
\label{tab:frame_counts}
\end{table}

We use the smaller versions of these datasets, \textit{NPR-small} and \textit{EB-NeRD-small}, since the core idea of this research is to analyze the impact of media frames in recommendations. These subsets provide sufficient samples to conduct and validate our experiments while maintaining computational efficiency. Table~\ref{tab:dataset_stats_adjusted} shows the distribution of articles, users, and impressions across the train, validation, and test splits used for experimentation. Note that the news categories in the NPR dataset are not topical; they primarily consist of abbreviations for Brazilian states. The distribution of news categories across datasets is shown in Table \ref{tab:npr_ebnerd_proportions_only}.
The overall distribution of frames across the two datasets is summarized in Table ~\ref{tab:frame_counts}. The observed differences in frame frequencies likely stem from the distinct origins and content of each corpus. Each dataset reflects the unique editorial focus, topical coverage, and national context of its source material. Additionally, variations in writing style across datasets may affect the performance of the frame classification model, potentially contributing to apparent shifts in frame distribution.

\begin{table}[!h]
\centering
\begin{adjustbox}{width=0.7\columnwidth, center}
\begin{tabular}{lccccccc}
\toprule
\textbf{Dataset} & \textbf{\#Articles} & \multicolumn{3}{c}{\textbf{Users}} & \multicolumn{3}{c}{\textbf{Impressions}} \\
\cmidrule(lr){3-5} \cmidrule(lr){6-8}
&  & \textbf{Train} & \textbf{Val} & \textbf{Test} & \textbf{Train} & \textbf{Val} & \textbf{Test} \\
\midrule
NPR-small & 65305 & 57617 & 9099 & 17888 & 64860 & 9267 & 18531 \\
EB-NeRD-small & 20738 & 14968 & 8811 & 15342 & 209598 & 23289 & 244647 \\
\bottomrule
\end{tabular}
\end{adjustbox}
\caption{Dataset statistics including number of articles, users, and impressions across train/val/test splits.}
\label{tab:dataset_stats_adjusted}
\end{table}

\section{Experimental Setup}

\begin{figure}[!h]
    \centering
    \begin{subfigure}[t]{0.8\textwidth}
        \centering
        \includegraphics[width=\linewidth]{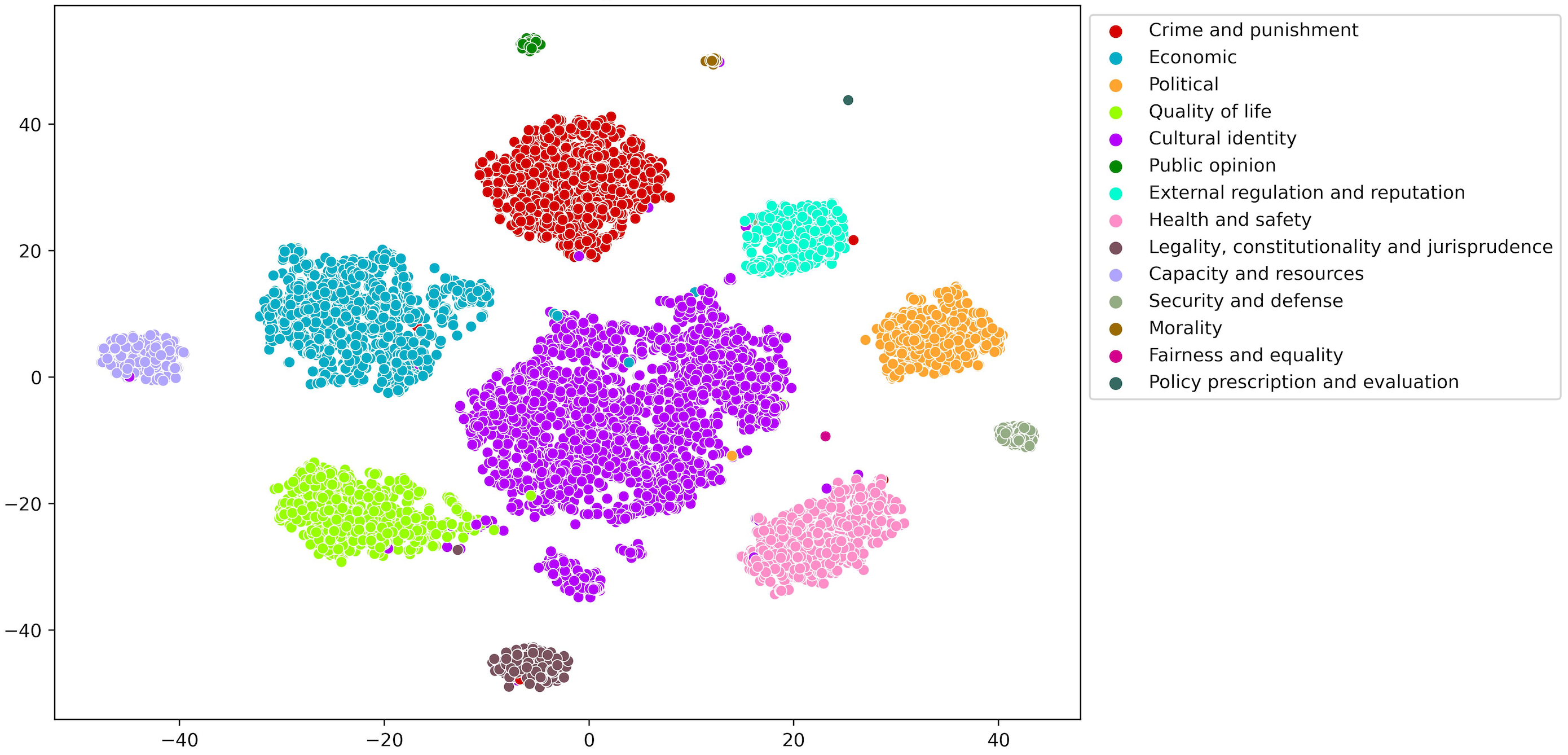}
        \caption{Frame-shaped embedding space on NPR test dataset.}
        \label{fig:npr-aspect}
    \end{subfigure}

    \vspace{1em}

    \begin{subfigure}[t]{0.8\textwidth}
        \centering
        \includegraphics[width=\linewidth]{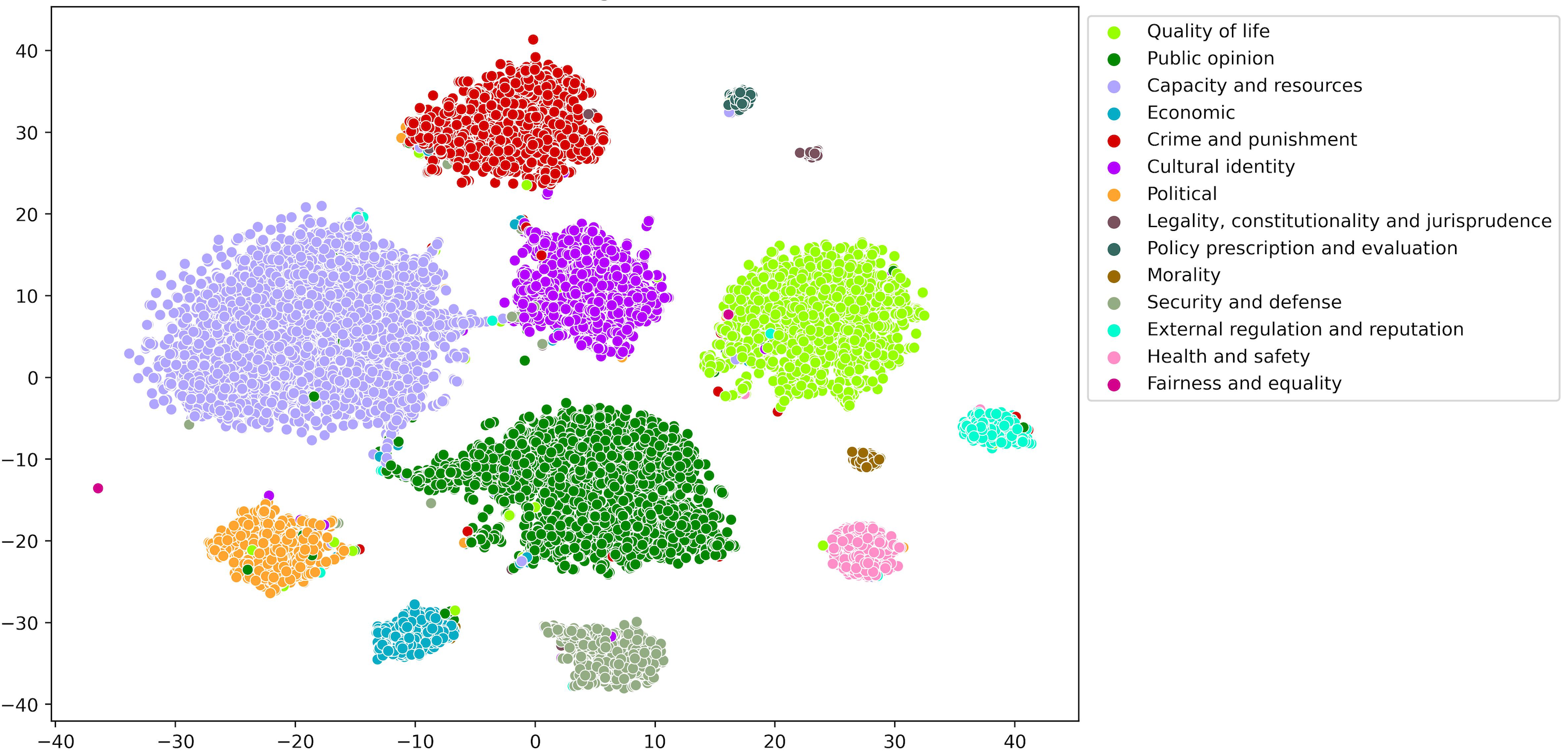}
        \caption{Frame-shaped embedding space on EB-NeRD test dataset.}
        \label{fig:ebnerd-aspect}
    \end{subfigure}
    \caption{t-SNE visualizations of news embeddings after training the frame-based aspect module.}
    \label{fig:aspect-comparison}
\end{figure}

\subsection{Training the Recommendation Systems}

We adopt training settings similar to those used in MANNeR. We use  \texttt{XLM-RoBERTa-base} in our experiments on both NPR and EB-NeRD, leveraging its multilingual capabilities to accommodate the cross-lingual nature of the datasets. The maximum user history length is set to 50 for both datasets. All models are trained using mixed precision and optimized with the Adam optimizer \citep{kingma2017adammethodstochasticoptimization}. For the frame-based aspect module, we use a learning rate of 1e-5 across both datasets. To address class imbalance, we sample 20 instances per class during the aspect module (Frame-Module) training. For the contrastive CR-Module, we adopt a 4:1 negative-to-positive sampling ratio and set the temperature parameter to 0.9 across all datasets. We train the CR-Module and baseline models for 5 epochs on NPR and 20 epochs on EB-NeRD, with a batch size of 8. The Frame-Module is trained for 100 epochs with a batch size of 60. Both modules employ early stopping with a patience of 3 epochs. All experiments are repeated five times with different random seeds, and we report the mean and standard deviation of evaluation metrics across runs. Training is conducted on an NVIDIA RTX A6000 GPU with 48 GB memory. Figure~\ref{fig:aspect-comparison} shows the embedding space after training the Frame-Module. The results show that the encoder groups well instances with the same frame closer together in both datasets.  
\section{Results and Discussion}

\begin{table}[!h]
\centering
\small

\begin{subtable}{\textwidth}
\centering
\resizebox{\linewidth}{!}{%
\begin{tabular}{c|cccccccc}
\toprule
$\lambda$ & AUC & MRR & @5 & @10 & Cal(C) & Cal(F) & Rep(F) & Act \\
\midrule
-1.0 & 66.75{\scriptsize$\pm$0.5} & 40.27{\scriptsize$\pm$0.3} & 45.12{\scriptsize$\pm$0.2} & 54.52{\scriptsize$\pm$0.4} & \textbf{87.47{\scriptsize$\pm$0.3}} & \textbf{78.45{\scriptsize$\pm$0.2}} & \textbf{49.77{\scriptsize$\pm$0.5}} & 58.45{\scriptsize$\pm$0.3} \\
-0.4 & 75.19{\scriptsize$\pm$0.3} & 50.07{\scriptsize$\pm$0.3} & 55.90{\scriptsize$\pm$0.4} & 62.28{\scriptsize$\pm$0.3} & 86.51{\scriptsize$\pm$0.3} & 75.67{\scriptsize$\pm$0.2} & 48.53{\scriptsize$\pm$0.3} & 58.01{\scriptsize$\pm$0.3} \\
-0.1 & 77.74{\scriptsize$\pm$0.3} & 51.29{\scriptsize$\pm$0.3} & 56.94{\scriptsize$\pm$0.4} & 63.10{\scriptsize$\pm$0.3} & 85.64{\scriptsize$\pm$0.3} & 70.15{\scriptsize$\pm$0.2} & 45.51{\scriptsize$\pm$0.4} & 58.46{\scriptsize$\pm$0.3} \\
0.0 & 78.08{\scriptsize$\pm$0.3} & 48.86{\scriptsize$\pm$0.3} & 57.86{\scriptsize$\pm$0.3} & 63.77{\scriptsize$\pm$0.3} & 85.40{\scriptsize$\pm$0.2} & 68.77{\scriptsize$\pm$0.2} & 45.14{\scriptsize$\pm$0.4} & 58.60{\scriptsize$\pm$0.3} \\
0.1 & \textbf{78.13{\scriptsize$\pm$0.4}} & \textbf{52.73{\scriptsize$\pm$0.3}} & \textbf{58.71{\scriptsize$\pm$0.4}} & \textbf{64.37{\scriptsize$\pm$0.4}} & 85.39{\scriptsize$\pm$0.3} & 68.40{\scriptsize$\pm$0.3} & 44.89{\scriptsize$\pm$0.5} & 58.61{\scriptsize$\pm$0.3} \\
0.4 & 77.00{\scriptsize$\pm$0.3} & 52.31{\scriptsize$\pm$0.3} & 58.28{\scriptsize$\pm$0.3} & 64.04{\scriptsize$\pm$0.3} & 85.24{\scriptsize$\pm$0.2} & 66.28{\scriptsize$\pm$0.3} & 43.88{\scriptsize$\pm$0.4} & 58.53{\scriptsize$\pm$0.3} \\
1.0 & 73.10{\scriptsize$\pm$0.4} & 48.56{\scriptsize$\pm$0.4} & 56.10{\scriptsize$\pm$0.3} & 60.07{\scriptsize$\pm$0.3} & 84.90{\scriptsize$\pm$0.3} & 63.81{\scriptsize$\pm$0.3} & 43.59{\scriptsize$\pm$0.5} & \textbf{58.70{\scriptsize$\pm$0.3}} \\
\bottomrule
\end{tabular}}
\caption{NPR}
\end{subtable}

\vspace{1em} % space between subtables

\begin{subtable}{\textwidth}
\centering
\resizebox{\linewidth}{!}{%
\begin{tabular}{c|cccccccc}
\toprule
$\lambda$ & AUC & MRR & @5 & @10 & Cal(C) & Cal(F) & Rep(F) & Act \\
\midrule
-1.0 & 56.31{\scriptsize$\pm$0.6} & 33.48{\scriptsize$\pm$0.2} & 39.26{\scriptsize$\pm$0.3} & 47.16{\scriptsize$\pm$0.6} & 74.42{\scriptsize$\pm$0.3} & 65.49{\scriptsize$\pm$0.4} & 71.06{\scriptsize$\pm$0.5} & \textbf{60.94{\scriptsize$\pm$0.3}} \\
-0.4 & 59.59{\scriptsize$\pm$0.2} & 35.89{\scriptsize$\pm$0.3} & 42.18{\scriptsize$\pm$0.4} & 49.49{\scriptsize$\pm$0.5} & \textbf{74.59{\scriptsize$\pm$0.3}} & \textbf{66.92{\scriptsize$\pm$0.3}} & 72.65{\scriptsize$\pm$0.4} & 59.16{\scriptsize$\pm$0.2} \\
-0.1 & 61.04{\scriptsize$\pm$0.4} & 36.08{\scriptsize$\pm$0.2} & 42.31{\scriptsize$\pm$0.3} & 49.61{\scriptsize$\pm$0.4} & 70.78{\scriptsize$\pm$0.3} & 62.87{\scriptsize$\pm$0.3} & 70.70{\scriptsize$\pm$0.2} & 59.40{\scriptsize$\pm$0.3} \\
0.0 & 61.44{\scriptsize$\pm$0.2} & 36.20{\scriptsize$\pm$0.3} & 43.70{\scriptsize$\pm$0.3} & 50.68{\scriptsize$\pm$0.5} & 70.67{\scriptsize$\pm$0.3} & 62.99{\scriptsize$\pm$0.3} & 70.95{\scriptsize$\pm$0.4} & 59.46{\scriptsize$\pm$0.2} \\
0.1 & 61.71{\scriptsize$\pm$0.3} & \textbf{38.26{\scriptsize$\pm$0.3}} & \textbf{44.75{\scriptsize$\pm$0.3}} & \textbf{51.53{\scriptsize$\pm$0.5}} & 70.64{\scriptsize$\pm$0.3} & 63.01{\scriptsize$\pm$0.3} & 70.99{\scriptsize$\pm$0.4} & 59.43{\scriptsize$\pm$0.3} \\
0.4 & \textbf{61.72{\scriptsize$\pm$0.2}} & \textbf{38.26{\scriptsize$\pm$0.3}} & \textbf{44.75{\scriptsize$\pm$0.2}} & \textbf{51.53{\scriptsize$\pm$0.4}} & 70.61{\scriptsize$\pm$0.3} & 63.72{\scriptsize$\pm$0.3} & 71.64{\scriptsize$\pm$0.5} & 59.54{\scriptsize$\pm$0.2} \\
1.0 & 60.48{\scriptsize$\pm$0.3} & 37.71{\scriptsize$\pm$0.3} & 44.00{\scriptsize$\pm$0.2} & 50.69{\scriptsize$\pm$0.4} & 71.11{\scriptsize$\pm$0.3} & 65.91{\scriptsize$\pm$0.4} & \textbf{73.69{\scriptsize$\pm$0.5}} & 60.34{\scriptsize$\pm$0.3} \\
\bottomrule
\end{tabular}}
\caption{EB-NeRD}
\end{subtable}

\caption{Performance across $\lambda$ for (a) NPR and (b) EB-NeRD. 
@5/@10 = nDCG@5/10; Cal(C/F) = Calibration on category/frame; Rep(F) = Representation on frames; Act = Activation. 
$\pm$ represents the standard deviation over runs with 5 different random seeds.}
\label{tab:npr_ebnerd_compact}
\end{table}

\begin{figure}[!h]
    \centering
    \begin{subfigure}[t]{1\textwidth}
        \centering
        \includegraphics[width=1\linewidth]{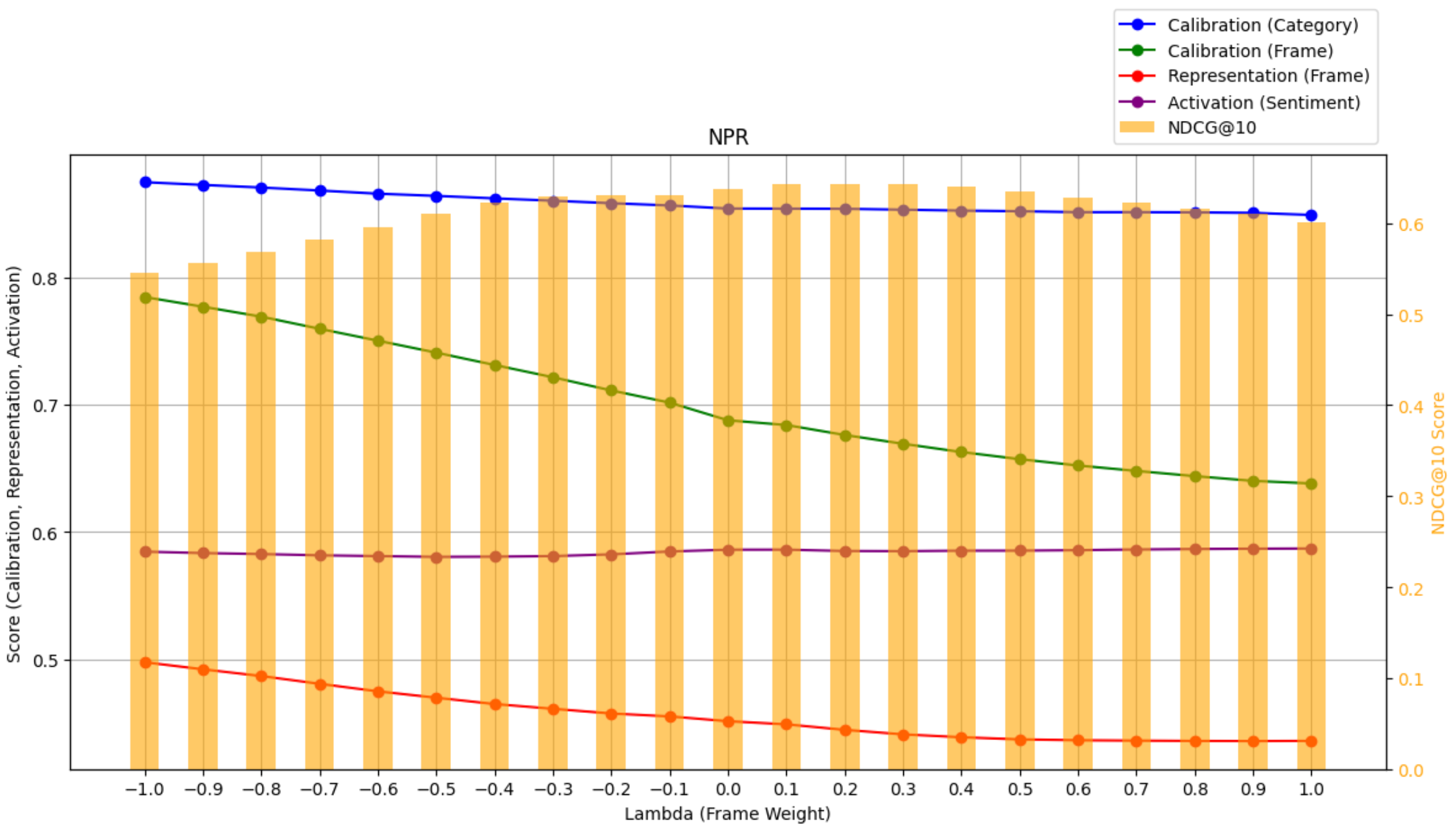}
        \caption{Frame aspect diversification for NPR.}
        \label{fig:npr-div}
    \end{subfigure}
    \hfill
    \begin{subfigure}[t]{1\textwidth}
        \centering
        \includegraphics[width=1\linewidth]{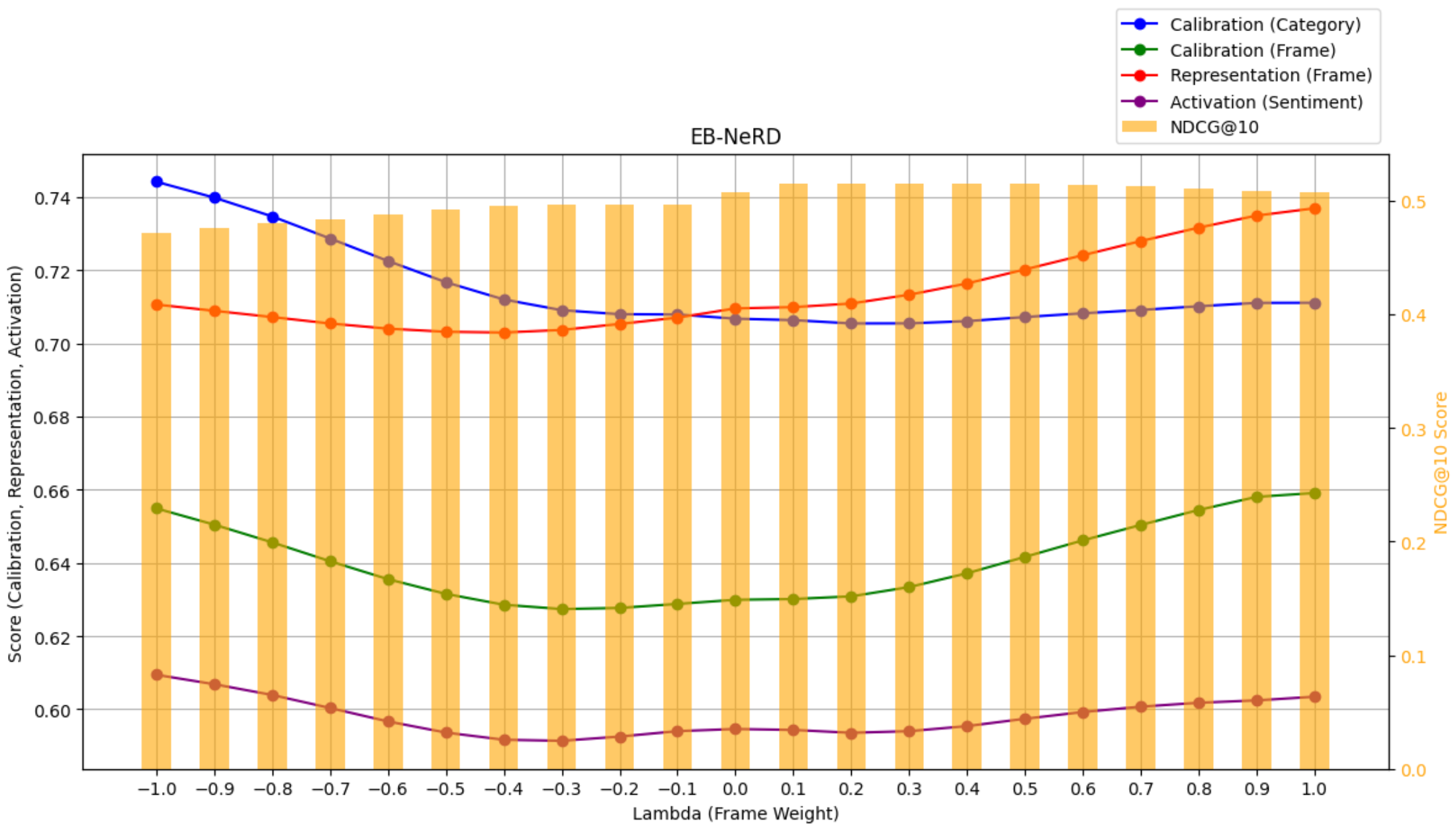}
        \caption{Frame aspect diversification for EB-NeRD.}
        \label{fig:ebnerd-div}
    \end{subfigure}
    \caption{Comparison of diversity-aware scores across EB-NeRD and NPR test datasets.}
    \label{fig:div-comparison}
\end{figure}

Table~\ref{tab:npr_ebnerd_compact} reports the mean $\pm$ standard deviation of evaluation metrics over 5 independent runs with different random seeds. To assess the effect of the diversification parameter $\lambda$, we systematically varied it from -1.0 to 1.0. Figure~\ref{fig:div-comparison} illustrates the resulting changes in nDCG@10 and the corresponding normative metric scores.

\paragraph{Impact on Personalization and User Satisfaction}

User satisfaction, measured by nDCG@5 and nDCG@10, peaks at moderate levels of personalization. In NPR, the highest scores occur at $\lambda = 0.1$, with nDCG@5 at $58.71 \pm 0.4$ and nDCG@10 at $64.37 \pm 0.4$. As personalization increases, performance declines—for instance, at $\lambda = 1.0$, nDCG@5 drops to $56.10 \pm 0.3$. A similar trend is observed with EB-NeRD, where nDCG@5 and nDCG@10 peak at $44.75 \pm 0.3$ and $51.53 \pm 0.5$ for $\lambda = 0.1$ and $\lambda = 0.4$, respectively.

These results indicate that excessive personalization ($\lambda = 1.0$) can introduce redundancy and overly narrow content selection. By relying too heavily on past preferences, recommendation systems risk producing repetitive outputs that miss users’ broader or evolving interests, ultimately reducing content diversity and perceived recommendation quality. This is a tendency also found in \citet{moeller-nr2025}.

\paragraph{Impact on Calibration}

Calibration on frames (Cal(F)) and categories (Cal(C)) is highest when the diversification parameter $\lambda$ is low and gradually decreases as $\lambda$ increases. In the NPR dataset, for instance, Cal(F) drops from 78.45$\pm$0.2 at $\lambda = -1.0$ to 63.81$\pm$0.3 at $\lambda = 1.0$. In the EB-NeRD dataset, on the other hand, the highest scores are reached already at $\lambda = -0.4$ with Cal(C) reaching 74.49$\pm$0.3 and Cal(F) equals 66.92$\pm$0.3 . This indicates that higher diversification (negative $\lambda$ values) leads to recommended articles whose frames and categories differ more significantly from the user's historical consumption. In other words, as diversification increases, the system prioritizes novelty over alignment with past user preferences, resulting in a greater mismatch between recommended content and user history, consequently increasing the number of perspectives the user has access to. 

Although categorical information was not explicitly encoded into the frame-based
diversification process, it nonetheless influenced categorical diversification. To analyze the impact, we examine the association between news category and frame class. Recognizing that statistical significance is common in large datasets, we calculate Cramér's V to assess the practical significance of the relationship. For the NPR dataset, the analysis yields a Cramér's V of 0.247, indicating a small to moderate association. For the EB-NeRD dataset, the Cramér's V is 0.332, suggesting a moderate association. These results confirm that while a predictable relationship exists between news category and frame, the effect sizes are not so large as to suggest that frames merely "shadow" news categories. A substantial portion of the variance in framing remains independent of the category.
Thus, promoting frame diversity can indirectly enhance category-level diversity as a side effect. This finding highlights a key trade-off: On the one hand, diversifying frames successfully enhances news category diversity, breaking up monolithic recommendation patterns. On the other hand, this side effect can conflict with the goal of providing varied perspectives on a stable news category, as it risks steering users away from their specific subject of interest. Further investigation is needed to understand whether adding news category into another A-module can keep the news categories the same while diversifying frames. This would satisfy the principle of diversifying perspectives on target news categories of interest to users.

\paragraph{Impact on Representation}
As diversification increases (i.e., lower $\lambda$ values), the frame representation score (Rep(F)) rises, indicating that the distribution of frames in the recommended content diverges more from that of the candidate pool. This reflects an increase in frame-level diversity, particularly in mitigating the dominance of over- or under-represented frames in the dataset. For example, in the NPR dataset, Rep(F) reaches its highest value  49.77$\pm$0.5 at $\lambda = -1.0$, and gradually declines to 43.59$\pm$0.5 at $\lambda = 1.0$. A different, though less pronounced, pattern is observed in EB-NeRD, where Rep(F) peaks at 73.69$\pm$0.5 for $\lambda = 1.0$, but remains consistently high across all $\lambda$ values due to the dataset’s inherent frame diversity.

\paragraph{Impact on Activation}
The Activation Score (\texttt{Act}), which measures deviation in sentiment tone, exhibits a non-linear response to the diversification parameter, $\lambda$. For the NPR dataset, the score initially decreases from $58.45 \pm 0.3$ at $\lambda = -1.0$ to a minimum of $58.01 \pm 0.3$ as $\lambda$ increases. The trend then reverses, with the \texttt{Act} rising to a peak of $58.70 \pm 0.3$ at higher $\lambda$ values. A similar U-shaped trend is observed for the EB-NeRD dataset: the \texttt{Act} is highest at $60.94 \pm 0.3$ ($\lambda = -1.0$), decreases to $59.16 \pm 0.2$ at $\lambda = -0.4$, and rises again to $60.34 \pm 0.3$ as personalization increases ($\lambda = 1.0$). These results suggest that both ends of the $\lambda$ spectrum, representing strong diversification and personalization, cause the sentiment in recommendations to shift further away from the average tone of the full content set.

\begin{figure}[!h]
    \centering
\includegraphics[width=1\linewidth]{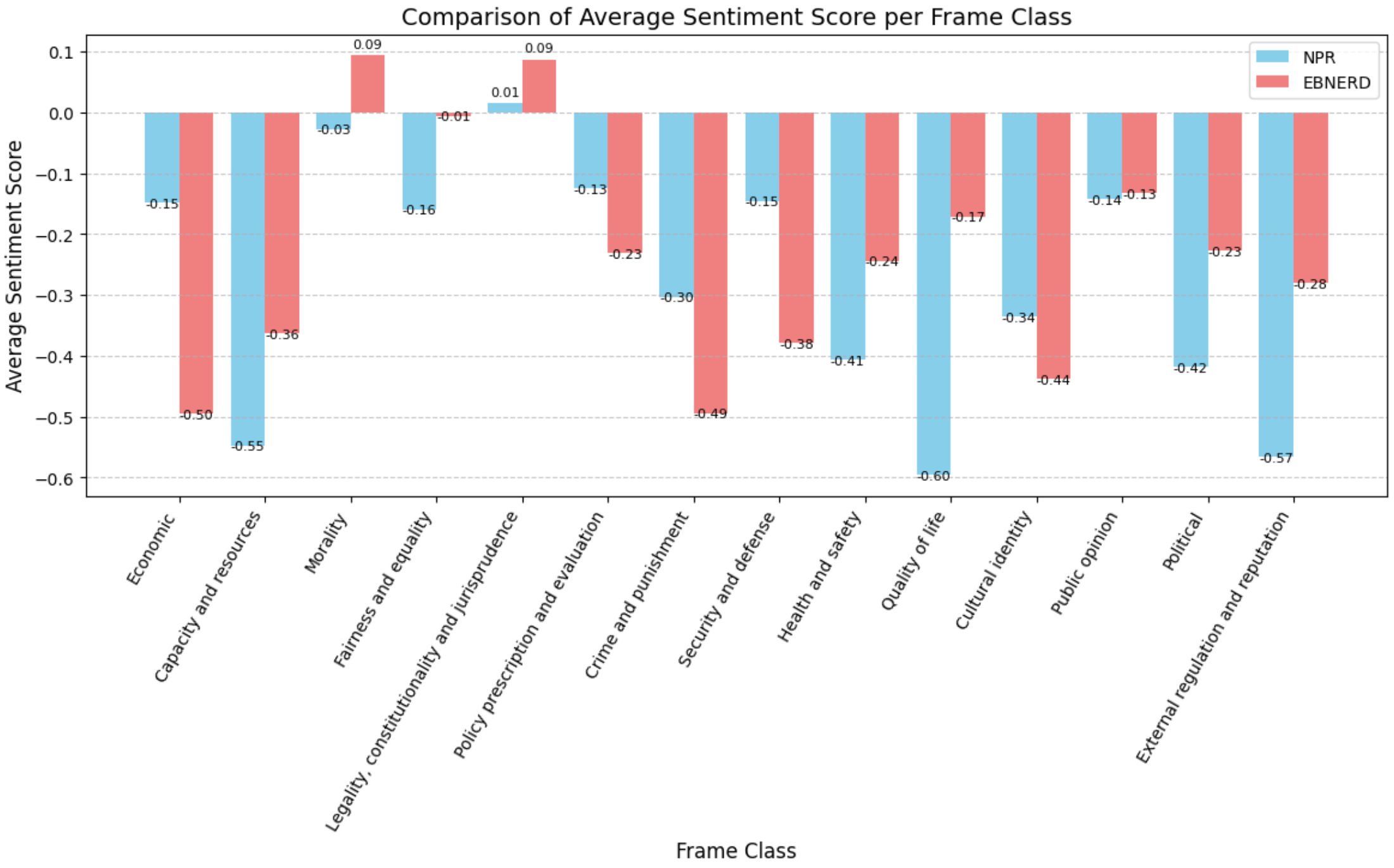}
    \caption{Distribution of average sentiment polarity across media frames.}
    \label{fig:sentiment-distribution}
\end{figure}

\begin{figure}[!h]
    \centering
    \begin{subfigure}[!h]{1\textwidth}
        \centering        \includegraphics[width=1\linewidth]{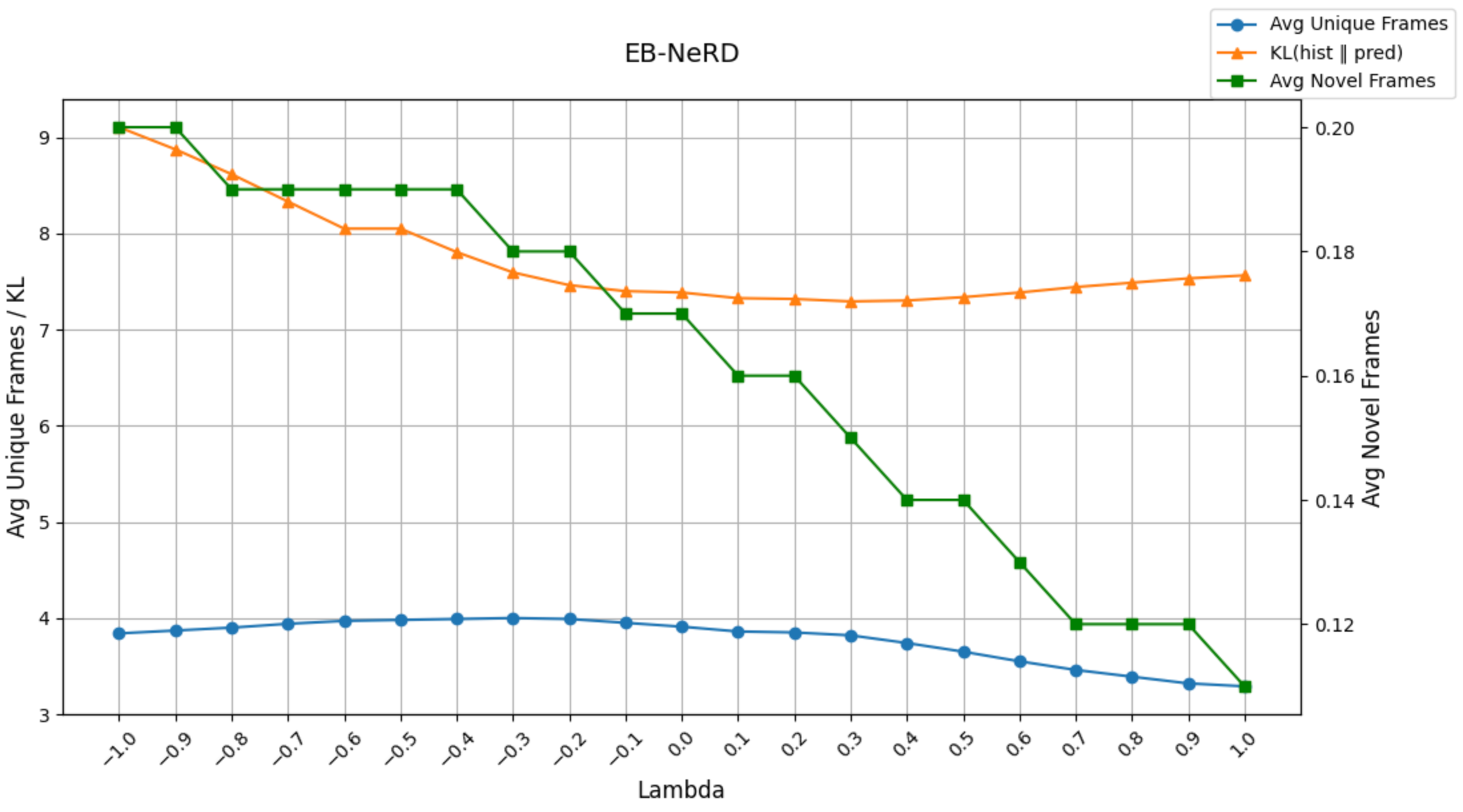}
        \caption{Impact on EB-NeRD test dataset.}
        \label{fig:ebnerd-novel}
    \end{subfigure}
    \hfill
    \begin{subfigure}[!h]{1\textwidth}
        \centering
    \includegraphics[width=1\linewidth]{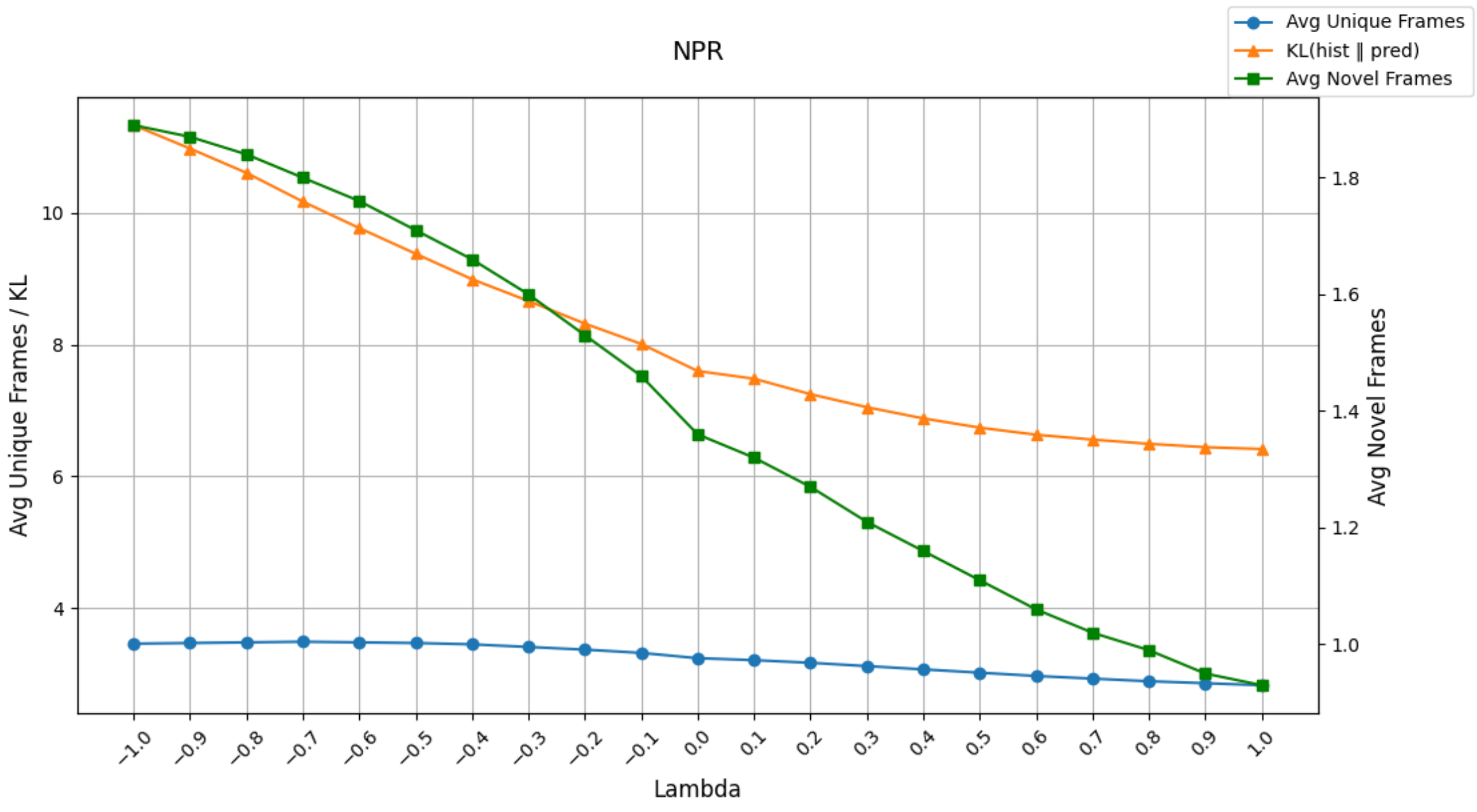}
        \caption{Impact on NPR test dataset.}
        \label{fig:npr-novel}
    \end{subfigure}
    \caption{Variation in novel frame exposure, KL divergence, and unique frames as a function of diversification level ($\lambda$).}
    \label{fig:novelty-side-by-side}
\end{figure}

Although sentiment was not an explicit feature encoded in our model, the diversification of framing strategies resulted in a certain extent of observable diversification in sentiment. To formally investigate this relationship, a one-way analysis of variance (ANOVA) was conducted to determine how much of the variation in sentiment scores could be explained by the different frame classes. The analysis revealed a statistically significant relationship in both datasets ($p < .001$). More importantly, an effect size analysis provided a measure of practical significance. For the NPR dataset, the frame class demonstrated a large effect on sentiment, accounting for 38.3\% of the total variance ($\eta^2 = 0.383$). Similarly, for the EB-NeRD dataset, the frame class had a large effect, explaining 20.0\% of the variance in sentiment scores ($\eta^2 = 0.200$). These findings show that frame selection has a strong influence on the sentiment of the text in both corpora. This strongly suggests that framing plays a crucial role in shaping the overall emotional tone, or polarity, of the content. Figure~\ref{fig:sentiment-distribution} illustrates the distribution of average sentiment polarity by media frame for both the NPR and EB-NeRD datasets. Notably, frames such as \textit{Morality}, \textit{Fairness and Equality}, and \textit{Legality} tend to exhibit slightly more positive sentiment polarity compared to frames like \textit{Economic}, \textit{Capacity and Resources}, and \textit{Crime and Punishment}. This indicates that emphasizing certain frames in the recommendation process can lead to a directional shift in the overall sentiment of the recommended content. However, the fact that they are on average negative across frames in both datasets explains why the \texttt{Act} does not vary considerably across setups.

Overall, the cost-benefit of the trade-off between accuracy and other objectives varies significantly between the two datasets. For the EB-NeRD dataset, the relationship is well-balanced; enhancing representation and calibration is achieved with a minimal cost of only 1-2 points to AUC. In contrast, the trade-off is considerably more pronounced for the NPR dataset, where similar enhancements incur a substantial drop of over 11 points in AUC.

\paragraph{Impact on Frame Novelty}

Figure~\ref{fig:novelty-side-by-side} illustrates the impact of the diversification parameter $\lambda$ ($-1.0$ to $1.0$) on the novelty of recommended frames for users of the NPR and EB-NeRD datasets. Three metrics are tracked: \textit{Average Unique Frames per User}, the mean number of distinct frames within each user's set of recommendations; \textit{KL Divergence} between a user's viewing history and their recommendations; and \textit{Average Novel Frames per User}, the mean count of recommended frames that the user has not previously encountered in their history. %\tani{Per user? Can you describe how you calculated that?}
 Across both datasets, increasing diversification ($\lambda$  = –1.0) leads to a clear rise in the number of novel frames shown to users and greater divergence from their historical preferences. This is evident in the peak values of novel frame exposure:  1.8 in NPR and 0.20 in EB-NeRD at $\lambda = -1.0$. KL divergence follows a similar trend, indicating that recommendations at higher diversification settings are less aligned with users’ prior interactions. A similar trend is observed in number of unique frames remaining in both datasets, showing only a slight decrease as $\lambda$ increases.

These results suggest that diversification effectively enhances the novelty and distinctiveness of recommended content across frames, although the extent of this effect is shaped by the underlying diversity in the user history within each dataset—more pronounced in NPR and more constrained in EB-NeRD. This difference can be attributed to the inherent frame diversity in users' reading histories: EB-NeRD users engage with a wider range of frames on average (11.8) compared to NPR users (3.8), leaving less room for further novelty through diversification. Indeed, as seen in the results above, the effect of the diversification in $\lambda$ is lower in the EB-NeRD dataset.

\section{Conclusion}

This work addresses the challenge of enhancing normative diversity in news recommendation by incorporating frame-level information in the recommendation system, as aspect of perspectives that is grounded in how news media construct meaning and shape public understanding. For that, we propose a frame-aware diversification approach that builds on the MANNeR framework to implement frame diversity. We evaluate our algorithm on the RADio* metrics to understand the effect of frames in the normative provisions of recommendation systems. 

Our experiments demonstrate that this approach effectively exposes users to novel frames that they have not previously encountered, resulting in increased diversity in recommended perspectives. Additionally, we find that framing implicitly captures sentiment and categorical variation, leading to improvements in normative diversity metrics such as calibration -- without the need for explicit modeling of these aspects. The increase in the diversity of news categories, a byproduct of frame-aware news recommendation, also enhances the diversity of content. 

We also observe a controllable trade-off between personalization and frame diversity. While prioritizing frame diversity improves normative aspects, it can result in a slight reduction in personalization metrics, such as accuracy and relevance. This trade-off is predictable and tunable, allowing system designers to strike a balanced configuration that maintains user relevance while promoting exposure to diverse frames. The inclusion of two datasets in our study enabled validation across different scenarios. We found that dataset composition can influence diversification outcomes; for example, greater heterogeneity in the EB-NeRD dataset makes differentiation more challenging but also reduces the risk of excessive personalization. These results highlight that the characteristics of the data should be carefully considered when tuning for diversity.

Together, these findings position media frames as a powerful and interpretable control lever for advancing normative diversity in news recommendation systems. Besides opening research directions for the academic community, our work has potential practical implications for media organizations. The adoption of frames on news platforms can support democratic diversification, fostering social and political engagement by the companies, while still allowing diversity tuning to preserve user satisfaction and interest. Moreover, users who are presented with a variety of viewpoints on social media are theoretically stimulated to participate (e.g., commenting, liking, sharing), thereby increasing interaction. To promote transparency and reproducibility for the wider community, we make the code available on GitHub [https://github.com/sourabhdattawad/framerec].

\section{Future Work}

This study employed the MANNeR framework to explore framing-based diversification. Future work can, however, investigate how other diversification methods or frameworks interact with framing. Our experiments were limited to smaller datasets (EB-NeRD-Small and NPR-Small); scaling to full-size versions could improve benchmarking and generalizability.

We also only encode the primary frame of each article, though many articles contain multiple frames. Extending the analysis to capture this complexity could help us understand its impact on diversification and how different framing combinations influence user understanding and engagement.

To address the side effect of frame diversity indirectly enhancing category-level diversity, we could extend the MANNeR framework by adding an additional category module alongside the existing frame module. This would allow us to diversify recommendations by frame while personalizing them by category. In this way, users are exposed to different frames without deviating from their categories of interest.

An important limitation of our approach is error propagation from the frame detection stage into the diversity analysis. Since the frame classifier achieves only moderate out-of-domain performance (F1 $\approx$ 0.48 on Portuguese data), misclassifications may distort both calibration and representation scores. As a result, observed differences in divergence metrics should be interpreted with caution. Future work could explore robustness analyses that explicitly quantify the impact of classifier errors on downstream diversity measurements.

While our current approach includes calibration, representation, and activation metrics, future work could expand the RADio* framework with new metrics -- namely fragmentation and alternative voices -- the former to examine how framing can tackle users’ isolation by reducing exposure to only narrow or isolated viewpoints that are not widely shared among users, and the latter by affecting the visibility of minority perspectives.

\section*{Acknowledgements}
We gratefully acknowledge funding from the German Federal Ministry of Research, Technology and Space (BMFTR) under the grant 01IS23072 for the Software Campus project MULTIVIEW.

%%
%% Define the bibliography file to be used
\bibliography{sample-ceur}

\end{document}